\documentclass[12pt]{revtex4}

\usepackage{amsmath}
\usepackage{rotating}
\usepackage{graphicx}
\usepackage{lscape}

\begin{document}

\noindent \large {\bf Comment on 'Energy partitioning schemes: a
dilemma' [I. Mayer, Faraday Discuss., (2007) 135, 439]}

\vskip 5mm

\normalsize

\noindent
Diego R. Alcoba$^{a}$, Roberto C. Bochicchio$^{a,*}$,
Alicia Torre$^{b}$, Luis Lain$^{b}$

\vskip 3mm

\noindent
$^{a}${\it Departamento de F\'{\i}sica, Facultad de
Ciencias Exactas y Naturales, Universidad de Buenos Aires and
Instituto de F\'{\i}sica de Buenos Aires, Consejo Nacional de
Investigaciones Cient\'{\i}ficas y T\'ecnicas. Ciudad Universitaria,
1428, Buenos Aires, Argentina}

\noindent $^{b}${\it Departamento de Qu\'{\i}mica F\'{\i}sica.
Facultad de Ciencia y Tecnolog\'{\i}a. Universidad del Pa\'{\i}s
Vasco. Apdo. 644 E-48080 Bilbao, Spain}

\vskip 3mm

\noindent \rule{155mm}{0.2mm}

\noindent {\bf Abstract} \vskip 2mm

\noindent A study of some decomposition schemes of the molecular
energy into one- and two-center contributions published in the above
mentioned journal highlights the importance of a 'dilemma' raised in
such decompositions. Even more, it has been recently assigned a
prominent role in the promotion energy mechanism. This critical
comment clarifies the validity of such a 'dilemma'.

\noindent \rule{155mm}{0.2mm} \vskip 5mm

Two decomposition schemes of the molecular energy into one- and
two-center contributions, $E = \sum_{A}\; E_{A}\; +\; \sum_{A<B}\;
E_{AB}$ have been reported in Refs. \cite{Mayer_dilemma, Mayer_cpl}.
They only differ in the treatment of the kinetic energy terms what
has been considered as a dilemma \cite{Mayer_dilemma}. One of them
(model 1) yields too large numbers (in absolute value) but presents
an expected distance behavior; the other (model 2) provides numbers
on the chemical scale but exhibits a counterintuitive distance
dependence. The aim of this report is to study the behavior of both
schemes in a wide interval of internuclear

\noindent
\rule{20mm}{0.4mm}\\
\noindent
{\footnotesize $^{*}$ Corresponding author. Fax: +54-11-45763357.\\
{\it E-mail address:} rboc@df.uba.ar (R.C. Bochicchio)}

\noindent
distances in order to check the existence of that dilemma.
The two partitioning schemes are \cite{Mayer_dilemma},

\[
E_{A}^{(1)}=\; \sum_{\mu \in A} \sum_{\nu}\; D_{\mu \nu} h_{\nu
\mu}^{A}\; + \frac{1}{2}\; \sum_{\mu, \rho \in A}
\sum_{\nu,\tau}\; \left( D_{\mu \nu} D_{\rho \tau}\; -\;
\frac{1}{2}\; D_{\rho \nu} D_{\mu \tau} \right) [\nu \tau|\mu \rho]
\]

\[
E_{AB}^{(1)}=\;\frac{Z_{A}Z_{B}}{R_{AB}}\; -\; \sum_{\mu \in A}
\sum_{\nu}\; D_{\mu \nu} \left<\nu \left| \frac{Z_{B}}{r_{B}}
\right| \mu \right>\; -\; \sum_{\mu \in B} \sum_{\nu}\; D_{\mu
\nu} \left<\nu \left| \frac{Z_{A}}{r_{A}} \right| \mu \right>\;
\]
\begin{equation}
+\; \sum_{\mu \in A} \sum_{\rho \in B}\; \sum_{\nu,\tau}\;
\left( D_{\mu \nu} D_{\rho \tau}\; -\; \frac{1}{2}\; D_{\rho \nu}
D_{\mu \tau} \right) [\nu \tau|\mu \rho]
\end{equation}

\vskip 5mm

\[
E_{A}^{(2)}=\; \sum_{\mu, \nu \in A}\; D_{\mu \nu} T_{\nu \mu}\;
-\; \sum_{\mu \in A} \sum_{\nu}\; D_{\mu \nu} \left<\nu \left|
\frac{Z_{A}}{r_{A}} \right| \mu \right>\;
\]
\[
+\; \frac{1}{2}\; \sum_{\mu, \rho \in A} \sum_{\nu,\tau}\;
\left( D_{\mu \nu} D_{\rho \tau}\; -\; \frac{1}{2}\; D_{\rho \nu}
D_{\mu \tau} \right) [\nu \tau|\mu \rho]
\]
\[
E_{AB}^{(2)}=\;\frac{Z_{A}Z_{B}}{R_{AB}}\; +\; 2\; \sum_{\mu \in
A, \nu \in B}\; D_{\mu \nu} T_{\nu \mu}\; -\; \sum_{\mu
\in A} \sum_{\nu}\; D_{\mu \nu} \left<\nu \left|
\frac{Z_{B}}{r_{B}} \right| \mu \right>\; -\; \sum_{\mu \in B}
\sum_{\nu}\; D_{\mu \nu} \left<\nu \left| \frac{Z_{A}}{r_{A}}
\right| \mu \right>\;
\]
\begin{equation}
+\; \sum_{\mu \in A} \sum_{\rho \in B}\; \sum_{\nu,\tau}\;
\left( D_{\mu \nu} D_{\rho \tau}\; -\; \frac{1}{2}\; D_{\rho \nu}
D_{\mu \tau} \right) [\nu \tau|\mu \rho]
\end{equation}

\vskip 5mm

\noindent
in which the superscripts $(1)$ and $(2)$ stand for models 1 and 2, respectively.
In these formulas $\mu,\nu,\rho,\tau,....$ are non-orthogonal atomic orbitals
which constitute the basis set. $h^{A}=T-\frac{Z_{A}}{r_{A}}$ is the one-electron
term of the Hamiltonian related with the atom $A$, being $T$ the kinetic energy
operator and $\frac{Z_{A}}{r_{A}}$ the electron-nucleus interaction; $Z_{A}$ is
the nuclear charge of atom A and $r_{A}$ the electron-nucleus distance. $D$ is
the usual first-order reduced density matrix and $[\nu \tau|\mu \rho]$ the two-electron
integrals in the $[12|12]$ convention. Notation $\mu \in A$ indicates that orbital
$\mu$ is centered on the atom $A$.

We have tested the behavior of both models studying the dissociation
of the ethane molecule (eclipsed in its singlet ground state) into
two methyl groups. The calculations have been performed using the
GAMESS package \cite{Gamess} at the same level of approximation
(restricted Hartree-Fock (RHF)) with the same basis set (6-31G**)
that were used for this system in Ref. \cite{Mayer_dilemma}. The
results are gathered in Fig. 1, where the distance dependence of the
two-center energy component $E_{\mathrm{CC}}$ is reported for the
two schemes. Both curves differ in the value of their energy minimum
and in its localization. The partitioning of the kinetic energy into
one- and two-center contributions (model 2) leads to a curve
appreciably softer than when this quantity is only treated as a
one-center feature (model 1). Model 1 predicts an unphysical value
$E_{\mathrm{CC}}= 447.95$ kcal/mol (0.7139 a.u.) at the minimum of
the curve which is very far from the chemical scale, while a more
reasonable value $E_{\mathrm{CC}}=129.57$ kcal/mol (0.2065 a.u.) is
attained within model 2, in agreement with the experimental binding
energy value 100.5 kcal/mol (0.1602 a.u.) \cite{BDE_handbook}. These
results show that the assignment of the kinetic energy only to
one-center contributions causes an abrupt lowering of the energy
$E_{\mathrm{CC}}$ which may be attributed, as the author claimed, to
a fast decay of the kinetic energy contributions in comparison with
those of electrostatic nature. The two curves have been fitted by a
Morse potential \cite{Morse} showing that the points in the
dissociation hypersurfaces follow an expected behavior in both
cases. Hence, we may argue that the counterintuitive distance
dependence of the model 2, which is the conclusion reported in Refs.
\cite{Mayer_dilemma,Mayer_cpl}, arises from calculations there
performed in a too short interval of distances CC situated in the
lowering part of the curve.

\vskip 5mm

\noindent {\bf Acknowledgements}: We acknowledge the financial
support of the Projects X017 (Universidad de Buenos Aires), PIP No.
11220090100061 (Consejo Nacional de Investigaciones Cient\'{\i}ficas
y T\'ecnicas, Rep\'ublica Argentina), Grant No. CTQ2009-07459/BQU
(the Spanish Ministry of Science and Innovation) and Grant No.
GIU09/43 (Universidad del Pais Vasco). We thank the Universidad del
Pais Vasco for allocation of computational resources.


\begin{figure}
\begin{center}
\includegraphics[scale=1.0,width=5.0in,height=5.0in]{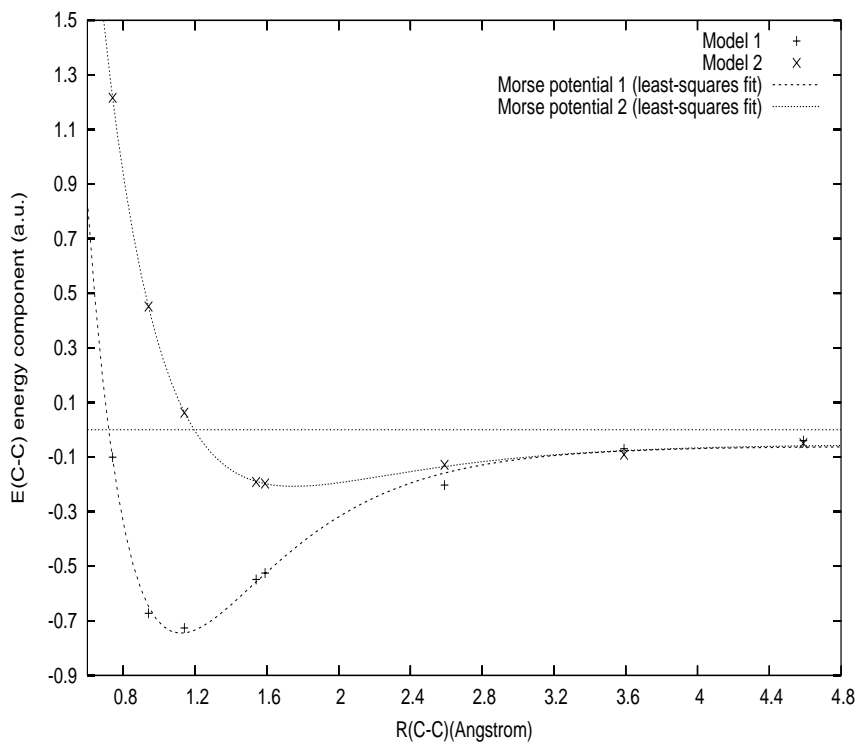}
\end{center}
\vskip -20mm \caption{Dissociation of the ethane molecule into two
methyl groups in eclipsed conformation at RHF level of approximation
using the 6-31G** basis set. The geometry of each $\mathrm{CH}_{3}$
group has been optimized at stretching.} \label{Fig. 1}
\end{figure}




\begin{references}

\bibitem{Mayer_dilemma} I. Mayer, Faraday Discuss. 135 (2007) 433.

\bibitem{Mayer_cpl} I. Mayer, Chem. Phys. Lett. 498 (2010) 366.

\bibitem{Gamess} M.W. Schmidt, K.K. Baldridge, J.A. Boatz, S.T. Elbert, M.S. Gordon,
J.H. Jensen, S. Koseki, N. Matsunaga, K.A. Nguyen, S. Su, T.L.
Windus, M. Dupuis, J.A. Montgomery, J. Comput. Chem. 14 (1993) 1347.

\bibitem{BDE_handbook} R. Luo, Comprehensive Handbook of Chemical Bond Energies, CRC Press,
Boca Raton, 2007.

\bibitem{Morse} K.I. Ramachandran, G. Deepa, K. Namboori, Computational Chemistry and
Molecular Modeling Principles and Applications, Springer-Verlag, Berlin, 2008.

\end{references}
\end{document}